\documentclass[sigconf]{acmart}
\usepackage{msc}
\usepackage{algorithmic}
\usepackage{graphicx}
\usepackage{acronym, comment, subcaption}
\usepackage{tabularx, pifont}
\usepackage[ruled,vlined,linesnumbered]{algorithm2e}
\usepackage{hyperref} 
\usepackage{mathtools}
\usepackage[colorinlistoftodos,textsize=small]{todonotes}
\usepackage{pdflscape}
\usepackage{lipsum}
\usepackage{multirow}
\usepackage{booktabs}

\usepackage{pgfplotstable}
\usepackage{colortbl}
\usepackage{xcolor}
\usepackage{etoolbox}

\usepackage{tikz}
\usetikzlibrary{chains,shapes.multipart}
\usetikzlibrary{shapes,calc}
\usetikzlibrary{automata,positioning}

\definecolor{customblue}{RGB}{0, 0, 0} 

\graphicspath{{Figures/}}
\acrodef{WLAN}{Wireless Local Area Network}
\acrodef{UN}{United Nations}
\acrodef{SDG}{Sustainable Development Goal}

\acrodef{MSE}{Mean Squared Error}
\acrodef{MAE}{Mean Absolute Error}
\acrodef{ML}{Machine Learning}
\acrodef{SVM}{Support Vector Machine}
\acrodef{DT}{Decision Tree}
\acrodef{RF}{Random Forest}
\acrodef{BSM}{Basic Safety Message}
\acrodef{DL}{Deep Learning}
\acrodef{RNN}{Recurrent Neural Network}
\acrodef{LSTM}{Long Short-Term Memory}
\acrodef{BiLSTM}{Bidirectional \ac{LSTM}}
\acrodef{CNN}{Convolutional Neural Network}
\acrodef{DDoS}{Distributed Denial-of-Service}
\acrodef{V2X}{Vehicle-to-Everything}
\acrodef{k-NN}{k-Nearest Neighbor}
\acrodef{KDE}{Kernel Density Estimation}
\acrodef{LLM}{Large Language Model}
\acrodef{MDS}{Misbehavior Detection System}
\acrodef{RSS}{Received Signal Strength}
\acrodef{VANET}{Vehicular Ad Hoc Network}
\acrodef{WEKA}{Waikato Environment for Knowledge Analysis}
\acrodef{DS}{Dempster-Shafer}
\acrodef{EL}{Ensemble Learning}
\acrodef{IDS}{Intrusion Detection System}
\acrodef{FLDS}{False Location Detection System}
\acrodef{Ens.RF}{Ensemble Random Forest}
\acrodef{CNB}{Complement Naive Bayes}
\acrodef{DTC}{Decision Tree Classifier}
\acrodef{GBC}{Gradient Boosting Classifier}
\acrodef{MDF}{Misbehavior Detection Framework}
\acrodef{MLP}{Multi-Layer Perceptron}
\acrodef{FL}{Federated Learning}
\acrodef{RL}{Reinforcement Learning}
\acrodef{DML}{Deep Multimodal Learning}
\acrodef{IoV}{Internet of Vehicle}
\acrodef{GRU}{Gated Recurrent Unit}
\acrodef{TPR}{True Positive Rate}
\acrodef{FPR}{False Positive Rate}

\acrodef{AAA}{Authentication, Authorization and Accounting}
\acrodef{ACC}{Adaptive Cruise Control}
\acrodef{ACL}{Access Control List}
\acrodef{AD}{Automated Driving}
\acrodef{ADAM}{Adaptive Moment Estimation}
\acrodef{ADAS}{Advanced Driver Assistance Systems}
\acrodef{AKI}{Accountable Key Infrastructure}
\acrodef{AoA}{Angle of Arrival}
\acrodef{API}{Application Programming Interface}
\acrodef{ASS}{Anonymity Set Size}
\acrodef{ATI}{Attack Time Impact}
\acrodef{AUC}{Area Under the Curve}
\acrodef{BSM}{Basic Safety Message}
\acrodef{BIC}{Bayesian Information Criterion}
\acrodef{BYOD}{Bring Your Own Device}
\acrodef{C2C-CC}{Car2Car Communication Consortium}
\acrodef{C2C}{Car-to-Car}
\acrodef{C2I}{Car-to-Infrastructure}
\acrodef{CA}{Certification Authority}
\acrodef{CAN}{Controller Area Network}
\acrodef{CHMM}{Continuous Hidden Markov Model}
\acrodef{CMIX}{Cryptographic Mix-Zone}
\acrodef{CN}{Common Name}
\acrodef{CVS}{Constant Vehicle Spacing}
\acrodef{CACC}{Cooperative Adaptive Cruise Control}
\acrodef{CAM}{Cooperative Awareness Message}
\acrodef{CAMP VSC3}{Crash Avoidance Metrics Partnership Vehicle Safety Consortium}
\acrodef{CAV}{Cooperative Aware Vehicle}
\acrodef{CDG}{Constant-Distance Gap}
\acrodef{CIA}{Confidentiality, Integrity and Availability}
\acrodef{CPU}{Central Processing Unit}
\acrodef{CRL}{Certificate Revocation List}
\acrodef{CTG}{Constant-Time Gap}
\acrodef{CTH}{Constant Time Headway}
\acrodef{CDN}{Content Delivery Network}
\acrodef{C-ITS}{Cooperative Intelligent Transport System}
\acrodef{COCA}{Cornell OnLine Certification Authority}
\acrodef{CSR}{Certificate Signing Request}
\acrodef{DAA}{Direct Anonymous Attestation}
\acrodef{DDoS}{Distributed Denial of Service}
\acrodef{DDH}{Decisional Diffie-Helman}
\acrodef{DENM}{Decentralized Environmental Notification Message}
\acrodef{DHMM}{Discrete Hidden Markov Model}
\acrodef{DHT}{Distributed Hash Table}
\acrodef{DL/ECIES}{Discrete Logarithm and Elliptic Curve Integrated Encryption Scheme}
\acrodef{DNN}{Deep Neural Network}
\acrodef{DoS}{Denial of Service}
\acrodef{DoT}{Department of Transportation}
\acrodef{DPA}{Data Protection Agency}
\acrodef{DSRC}{Dedicated Short Range Communication}
\acrodef{DSS}{Digital Signature Standard}
\acrodef{ECU}{Electronic Control Unit}
\acrodef{EDR}{Event Data Recorder}
\acrodef{ETSI}{European Telecommunications Standards Institute}
\acrodef{ECDSA}{Elliptic Curve Digital Signature Algorithm}
\acrodef{ECC}{Elliptic Curve Cryptography}
\acrodef{EVITA}{E-safety Vehicle Intrusion protected Applications}
\acrodef{FOT}{Field Operational Test}
\acrodef{FPGA}{Field-Programmable Gate Array}
\acrodef{GPA}{Global Passive Adversary}
\acrodef{GPS}{Global Positioning System}
\acrodef{GN}{GeoNetworking}
\acrodef{GS-VLR}{Group Signatures with Verifier Local Revocation}
\acrodef{GS}{Group Signatures}
\acrodef{GM}{Group Manager}
\acrodef{GBA}{Generic Bootstrapping Architecture}
\acrodef{GNSS}{Global Navigation Satellite System}
\acrodef{GUI}{Graphic User Interface}
\acrodef{HMM}{Hidden Markov Model}
\acrodef{GMM}{Gaussian Mixture Model}
\acrodef{GMMHMM}{Gaussian Mixture Model Hidden Markov Model}
\acrodef{HSM}{Hardware Security Module}
\acrodef{HTTP}{Hypertext Transfer Protocol}
\acrodef{IEEE}{Institute of Electrical and Electronics Engineers}
\acrodef{IID}{Independent and Identically Distributed}
\acrodef{IETF}{Internet Engineering Task Force}
\acrodef{IoT}{Internet of Things}
\acrodef{ITS}{Intelligent Transport Systems}
\acrodef{IT}{Information Technologies}
\acrodef{IVN}{In-Vehicle Network}
\acrodef{IMSI}{International Mobile Subscriber Identity}
\acrodef{IMEI}{International Mobile Station Equipment Identity}
\acrodef{IdP}{Identity Provider}
\acrodef{IDS}{Intrusion Detection System}
\acrodef{ISP}{Internet Service Provider}
\acrodef{LEA}{Law Enforcement Agency}
\acrodef{LCPP}{Lightweight Conditional Privacy Preservation}
\acrodef{LLR}{Log Likelihood Ratio}
\acrodef{LTC}{Long Term Certificate}
\acrodef{LTCA}{Long Term \acs{CA}}
\acrodef{MDS}{Misbehavior Detection Scheme}
\acrodef{H-LTCA}{Home-LTCA}
\acrodef{F-LTCA}{Foreign-LTCA}
\acrodef{LDAP}{Lightweight Directory Access Protocol}
\acrodef{LBS}{Location Based Service}
\acrodef{LSTM}{Long Short-Term Memory}
\acrodef{LTE}{Long Term Evolution}
\acrodef{LuST}{Luxembourg SUMO Traffic}
\acrodef{MA}{Misbehavior Authority}
\acrodef{MCP}{Maneuver Coordination Protocol}
\acrodef{MCM}{Maneuver Coordination Message}
\acrodef{MCS}{Maneuver Coordination Service}
\acrodef{MAC}{Media Access Control}
\acrodef{MCA}{Message \ac{CA}}
\acrodef{MEA}{Misbehavior Evaluation Authority}
\acrodef{MRM}{Minimum Risk Maneuver}
\acrodef{NN}{Neural Network}
\acrodef{NLP}{Natural Language Processing}
\acrodef{NTP}{Network Time Protocol}
\acrodef{OBU}{On-Board Unit}
\acrodef{OEM}{Original Equipment Manufacturer}
\acrodef{OCSP}{Online Certificate Status Protocol}
\acrodef{PAMPOS}{PAMPOS}
\acrodef{PCA}{Pseudonym \acs{CA}}
\acrodef{PDP}{Policy Decision Point}
\acrodef{PEP}{Policy Enforcement Point}
\acrodef{PIR}{Private Information Retrieval}
\acrodef{PKC}{Public Key Cryptography}
\acrodef{PKCS}{Public Key Cryptosystem}
\acrodef{PKI}{Public-Key Infrastructure}
\acrodef{PRECIOSA}{Privacy Enabled Capability in Co-operative Systems and Safety Applications}
\acrodef{PRESERVE}{Preparing Secure Vehicle-to-X Communication Systems}
\acrodef{PRIME}{Platoon Restructuring for Incident Mitigation and Exclusion}
\acrodef{P2P}{peer-to-peer}
\acrodef{PS}{Participatory Sensing}
\acrodef{RA}{Resolution Authority}
\acrodef{RAM}{Random Access Memory}
\acrodef{REST}{Representational State Transfer}
\acrodef{RBAC}{Role Based Access Control}
\acrodef{RCA}{Root \acs{CA}}
\acrodef{RSU}{Roadside Unit}
\acrodef{RHyTHM}{RHyTHM}
\acrodef{SAML}{Security Assertion Markup Language}
\acrodef{SAS}{Sample Aggregation Service}
\acrodef{SCMS}{Security Credential Management System}
\acrodef{SCORE@F}{Système COopératif Routier Expérimental Français}
\acrodef{SDSI}{Simple Distributed Security Infrastructure}
\acrodef{SRAAC}{Secure Revocable Anonymous Authenticated Inter-Vehicle Communication}
\acrodef{SeVeCom}{Secure Vehicle Communication}
\acrodef{SAE}{Society of Automotive Engineers}
\acrodef{SIT}{Sichere Informationstechnologie}
\acrodef{SLC}{Short-Lived Certificate}
\acrodef{SoA}{Service-oriented-Approach}
\acrodef{SIFS}{Short Inter Frame Space}
\acrodef{SSO}{Single-Sign-On}
\acrodef{SSL}{Secure Sockets Layer}
\acrodef{SOAP}{Simple Object Access Protocol}
\acrodef{STRP}{Spatial Time Reservation Procedure}
\acrodef{SVM}{Support Vector Machine}
\acrodef{TA}{Transition Area}
\acrodef{TACK}{Temporary Anonymous Certified Key}
\acrodef{TFLite}{TensorFlow Lite}
\acrodef{TS}{Task Service}
\acrodef{TLS}{Transport Layer Security}
\acrodef{ToC}{Transition of Control}
\acrodef{TPM}{Trusted Platform Module}
\acrodef{TTP}{Trusted Third Party}
\acrodef{TVR}{Ticket Validation Repository}
\acrodef{URI}{Uniform Resource Identifier}
\acrodef{UML}{Unified Modeling Language}
\acrodef{VANET}{Vehicular Ad-hoc Network}
\acrodef{V2I}{Vehicle-to-Infrastructure}
\acrodef{V2V}{Vehicle-to-Vehicle}
\acrodef{V2X}{Vehicle-to-Everything}
\acrodef{VC}{Vehicular Communication}
\acrodef{VM}{Virtual Machine}
\acrodef{VSS}{\ac{VC} Security Subsystem}
\acrodef{WAVE}{Wireless Access in Vehicular Environments}
\acrodef{WSDL}{Web Services Discovery Language}
\acrodef{W3C}{World Wide Web Consortium}
\acrodef{V}{Vehicle}
\acrodef{VANET}{Vehicular Ad-hoc Network}
\acrodef{VLR}{Verifier-Local Revocation}
\acrodef{VPKI}{Vehicular Public-Key Infrastructure}
\acrodef{VM}{Virtual Machine}
\acrodef{WS}{Web Service}
\acrodef{WoT}{Web of Trust}
\acrodef{WSACA}{\ac{WAVE} Service Advertisement \ac{CA}}
\acrodef{XML}{Extensible Markup Language}
\acrodef{XACML}{eXtensible Access Control Markup Language}
\acrodef{ZKP}{Zero Knowledge Proof}
\acrodef{3G}{3rd Generation}
\acrodef{ROC}{Receiver Operating Characteristic}
\acrodef{TP}{True Positive}
\acrodef{FP}{False Positive}
\acrodef{TN}{True Negative}
\acrodef{FN}{False Negative}

\acrodef{VAE}{Variational Autoencoder}
\acrodef{LR}{Learning Rate}

\copyrightyear{2026}
\acmYear{2026}
\setcopyright{cc}
\setcctype{by-nc-nd}
\acmConference[WiseML '26]{Proceedings of the 2026 ACM Workshop on Wireless Security and Machine Learning }{June 30-July 03, 2026}{Saarbrücken, Germany}
\acmBooktitle{Proceedings of the 2026 ACM Workshop on Wireless Security and Machine Learning (WiseML '26), June 30-July 03, 2026, Saarbrücken, Germany}
\acmDOI{10.1145/3811880.3815104}
\acmISBN{979-8-4007-2705-4/2026/06}

\begin{document}

\title[PAMPOS: Trajectory Prediction for Attack-Agnostic Misbehavior Detection in V2X Networks]{PAMPOS: Causal Transformer-based Trajectory Prediction for Attack-Agnostic Misbehavior Detection in V2X Networks}

\author{Konstantinos Kalogiannis}
\authornote{Equally contributing authors.}
\orcid{0000-0002-4656-2565}
\affiliation{%
  \institution{Networked Systems Security Group\\KTH Royal Institute of Technology}
  \city{Stockholm}
  \country{Sweden}
}
\email{konkal@kth.se}

\author{Ahmed Mohamed Hussain}
\authornotemark[1]
\orcid{0000-0003-4732-9543}
\affiliation{%
  \institution{Networked Systems Security Group\\KTH Royal Institute of Technology}
  \city{Stockholm}
  \country{Sweden}
}
\email{ahmhus@kth.se}

\author{Panos Papadimitratos}
\orcid{0000-0002-3267-5374}
\affiliation{%
  \institution{Networked Systems Security Group\\KTH Royal Institute of Technology}
  \city{Stockholm}
  \country{Sweden}
}
\email{papadim@kth.se}

\renewcommand{\shortauthors}{K. Kalogiannis, A. M. Hussain, and P. Papadimitratos}

\begin{abstract}
Misbehavior detection in \ac{V2X} networks is a second line of defense against insider falsification attacks that cryptographic mechanisms alone cannot address. Existing learning-based \acp{MDS} are supervised, requiring labeled attack samples at training time, thus failing to counter unseen falsification attacks. We present \acs{PAMPOS}, a causal transformer-decoder trained on benign VeReMi++ trajectories to learn normal mobility patterns. At inference time, misbehavior is identified as a deviation from the model's next-step kinematic predictions using a top-$K$ normalized anomaly scoring mechanism that localizes falsification to specific kinematic features, without requiring attack-labeled training data. We evaluate \acs{PAMPOS} across all 19 attack types in VeReMi++ under rush-hour and afternoon scenarios, achieving \ac{AUC} values of up to 0.98 and F1-scores of up to 0.95 for most attack categories.
\end{abstract}



\maketitle

\section{Introduction}
\label{sec:introduction}

{\color{customblue}\acp{VANET} and \acp{C-ITS} have emerged as key enablers of road safety, traffic efficiency, and situational awareness, relying on the continuous exchange of \acp{CAM} among \ac{V2X}-enabled vehicles. These messages contain kinematic information, including position, speed, acceleration, and heading, and serve as the foundation of cooperative perception in modern vehicular environments. The integrity of this shared traffic picture is therefore paramount: a vehicle acting on falsified \acp{CAM} can make incorrect and potentially dangerous driving decisions, such as emergency braking or unsafe lane changes.

Standards, such as the IEEE 1609.2 WG~\cite{IEEEWAVE2022} and~\ac{ETSI}~\cite{ETSI-TS-102-940}, secure \ac{VC} systems. Security architectures for \ac{VC} systems~\cite{PapadimitratosBHSFRMKKH2008} provide security and privacy protection based on vehicular \ac{PKI}, while preventing misuse of credentials~\cite{khodaei2018Secmace}. However, they offer no protection against insider threats: compromised vehicles, provisioned with credentials, that deliberately transmit falsified kinematic data. Such falsification attacks can be particularly dangerous until the attacker is revoked~\cite{KhodaeiP:J:2021b}, as their effects on surrounding vehicles can be immediate and severe, especially during dynamic scenarios such as lane changes or maneuvers~\cite {kalogiannis2023vulnerability, kalogiannis2022attack}. This motivates the need for \acp{MDS} that detect falsified data and trigger revocation~\cite{PapadimitratosBHSFRMKKH2008, RayaPapadiGH:C:2008}.

Traditional data-centric \acp{MDS} rely on plausibility checks and consistency thresholds, while recent \ac{ML} solutions, including \ac{LSTM}, \ac{CNN}, and transformer-based architectures, have demonstrated promising results on the VeReMi and VeReMi++ datasets~\cite{boualouache2023survey, hsu2021deep, gyawali2020machine, ercan2022misbehavior} and platooning benchmarks~\cite{li2025attentionguard, kalogiannis2025attention}. However, these approaches operate in a supervised manner: a detector trained on a fixed set of known attacks cannot reliably detect falsification strategies that deviate from its training distribution.

On the other hand, (semi-)unsupervised solutions~\cite{alladi2021deepadv, liu2023svmdformer, huang2025ultraadv, campos2025vae} reduce this dependency on labeled attack data by learning the statistical structure of normal behavior and flagging deviations at inference time. However, SVMDformer~\cite{liu2023svmdformer} still requires a small number of labeled samples for the hardest attack class. Campos et al.~\cite{campos2025vae} operate in a federated setting restricted to binary detection on the original VeReMi dataset, and all four approaches are limited to binary detection, leaving multi-class unsupervised misbehavior detection across heterogeneous attack taxonomies an open problem.

In this paper, we propose \acs{PAMPOS}, a causal transformer-based framework that reframes misbehavior detection as an unsupervised anomaly detection problem. It is trained exclusively on benign vehicular trajectories from VeReMi++~\cite{kamel2020veremi}, learning the predictive structure of normal mobility without exposure to any attack data. Deployed on a vehicle, at inference time, \acs{PAMPOS} identifies misbehavior as a statistically significant deviation from the model's next-step kinematic prediction, requiring no attack labels and generalizing naturally to unseen attack variants.

\textbf{Contributions.} (i) An unsupervised \ac{MDS} framework that requires no attack-labeled training data, enabling detection of novel attacks not seen during training, evaluated across all 19 attack types in VeReMi++. (ii) A causal transformer-decoder architecture trained on benign-only trajectories (i.e., attack-agnostic) with a top-$K$ normalized scoring mechanism that localizes anomalous behavior to specific kinematic features, informing on the performed attack family. (iii) An analysis of the limits of prediction-error-based detection, identifying constant position offset (A2) and eventual stop (A9) as open challenges due to their similarity to benign behavior.

\textbf{Paper Organization.} Sec.~\ref{sec:background} reviews related work on misbehavior detection datasets, \ac{ML} and \ac{DL} approaches for \acp{VANET}, and transformer-based modeling. Sec.~\ref{sec:adversary-model} presents the system and threat model. Sec.~\ref{sec:solution} describes the \acs{PAMPOS} framework, including data preprocessing, model architecture, and anomaly scoring. Sec.~\ref{sec:performance} presents the performance evaluation, discusses cross-scenario generalization, and inference efficiency. Finally, Sec.~\ref{sec:conclusion} concludes with key findings.}

\section{Related Work}
\label{sec:background}

\textbf{Misbehavior Detection Datasets.} Van Der Heijden et al.~\cite{van2018veremi} introduced VeReMi, the first publicly available dataset for the comparable evaluation of \acp{MDS} in \acp{VANET}, providing labeled \acp{CAM} across five position-falsification attack types and 3 attacker-density ratios. Kamel et al.~\cite{kamel2020f2md} proposed F$^{\text{2}}$MD, an open-source \ac{C-ITS} simulation framework built on OMNeT++ and VEINS for implementing attacks, evaluating detectors, and generating labeled datasets. Building on this, Kamel et al.~\cite{kamel2020veremi} generated VeReMi++, extending the original dataset with a realistic sensor-error model and approximately twenty attack variations, encompassing \ac{DoS}, data replay, Sybil, and falsification attacks across 24-hour simulation periods at diverse traffic densities. VeReMi++ serves as the primary evaluation dataset in this work.

\textbf{Misbehavior Detection Approaches.} Classical \ac{ML} techniques have demonstrated considerable efficacy in detecting anomalous vehicular behavior~\cite{boualouache2023survey}. F$^{\text{2}}$MD~\cite{kamel2020f2md} benchmarked \ac{SVM}, \ac{MLP}, and \ac{LSTM} classifiers, with \ac{LSTM} achieving the highest accuracy, confirming the advantage of sequential modeling for kinematic misbehavior detection. Gyawali et al.~\cite{gyawali2020machine} combined \ac{ML} with Dempster-Shafer theory for reputation-based detection, while~\cite{ercan2022misbehavior} used \ac{AoA} and signal-strength features with \ac{k-NN} and \ac{RF} to detect position falsification in \ac{C-ITS}.

\ac{DL} methods have further improved generalization: Liu et al.~\cite{liu2022misbehavior} reduced \ac{FP} rates using \ac{LSTM}; Hsu et al.~\cite{hsu2021deep} achieved 0.95 accuracy with a hybrid \ac{CNN}-\ac{LSTM} on VeReMi++; Alladi et al.~\cite{alladi2021deepadv} reported 0.98 accuracy with an F1-score of 0.97 across all VeReMi++ attack categories; and Youness et al.~\cite{youness2025vemisnet} achieved 0.99 accuracy with an F1-score of 0.99 using a \ac{BiLSTM} with domain-informed spatiotemporal features. All of these, however, require labeled attack samples during training, failing to generalize to novel attacks absent from the training distribution.

Giuliari et al.~\cite{giuliari2021transformer} studied the viability of transformer encoders for mobility modeling, outperforming \ac{LSTM} baselines on pedestrian forecasting. In \acp{VANET}, Li et al.~\cite{li2025attentionguard} proposed AttentionGuard, a multi-head transformer-encoder for platoon misbehavior detection, achieving an F1 of 0.95 and an \ac{AUC} of 0.96 with 100$ms$ decision latency. Kalogiannis et al.~\cite{kalogiannis2025attention} extended this with AIMformer, incorporating vehicle-specific temporal positional encoding and a precision-focused loss function to mitigate \acp{FP} under class imbalance, with further optimization for edge inference via TFLite, ONNX, and TensorRT. Both systems operate as supervised binary classifiers evaluated on platooning-specific datasets.

\textbf{Comparison with Existing Work.} \acs{PAMPOS} departs from the supervised paradigm adopted by the works above in two ways. Rather than training on labeled attack examples, it learns the predictive structure of normal mobility from benign-only trajectories, identifying misbehavior at inference time as a deviation from this learned prior without requiring any attack-labeled data. The closest works in this direction are DeepADV~\cite{alladi2021deepadv}, SVMDformer~\cite{liu2023svmdformer}, UltraADV~\cite{huang2025ultraadv}, and Campos et al.~\cite{campos2025vae}, yet a direct numerical comparison with these methods is not feasible given their incompatible experimental setup. SVMDformer deploys to \acp{RSU} rather than on-vehicle, requiring a full 200-message sequence before scoring, roughly 20 seconds at a typical \ac{CAM} rate, whereas \acs{PAMPOS} scores after just 10 messages. 

UltraADV similarly deploys to \acp{RSU} and selects its anomaly threshold using a test set containing artificially introduced attacks, making it semi-supervised rather than fully unsupervised; it also prioritizes deployment efficiency through knowledge distillation but does not address the interpretability of attack types. Campos et al. combine \acp{VAE} and \acp{GMM} within a \ac{FL} framework on VeReMi under non-\ac{IID} conditions, in a setting that is incomparable to centralized detection. 

All four are further limited to binary detection, whereas \acs{PAMPOS} addresses multi-class unsupervised detection across the full 19-attack-type VeReMi++ benchmark without federated overhead. Its top-$K$ feature attribution mechanism also localizes anomalies to specific kinematic dimensions, giving \ac{MA} actionable attack-type information beyond a binary flag, and \acs{PAMPOS} demonstrates cross-scenario generalization between rush-hour and afternoon traffic, a property none of these works evaluate. Furthermore, unlike AttentionGuard~\cite{li2025attentionguard} and AIMformer~\cite{kalogiannis2025attention}, which target small homogeneous platooning formations, \acs{PAMPOS} operates across the broader, more heterogeneous \ac{V2X} setting of VeReMi++, encompassing diverse sender populations and 19 attack types.
\section{System and Threat Model}
\label{sec:adversary-model}
We consider a \ac{V2X}-enabled urban vehicular network in which vehicles periodically broadcast \acp{CAM} containing their reported kinematic state, namely position, speed, acceleration, and heading, to their neighborhood. Vehicles receiving these messages, from surrounding senders, evaluate each sender's trajectory over time. We assume a threat model applicable to~\acp{VC}~\cite{PapadimitratosBHSFRMKKH2008}, with vehicles possessing valid credentials that allow them to broadcast \acp{CAM}~\cite{khodaei2018Secmace}. \acs{PAMPOS} serves as a second line of defense, running on each vehicle and detecting insider threats posed by such compromised vehicles.

Concretely, attackers traveling on the road can manipulate \acp{CAM} at multiple levels, spanning three categories~\cite{kamel2020veremi}: \textit{position and speed falsification}, where reported values are manipulated via constant offsets or random perturbations (A1--A8); \textit{behavioral attacks}, where the reported trajectory follows a physically plausible but malicious pattern, such as eventual stop or disruptive mobility (A9--A10); and \textit{protocol-level attacks}, including data replay, delayed message injection, and \ac{DoS} flooding, with Sybil variants of each (A11--A19). We make no assumption about the attacker's knowledge of the detection mechanism, including attackers whose falsified trajectories remain physically consistent with normal driving---such as constant position offset (A2) and eventual stop (A9)---which represent a known open challenge discussed in Sec.~\ref{sec:performance}.

\section{Proposed Framework -- PAMPOS}
\label{sec:solution}

\begin{figure*}[!ht]
    \centering
    \includegraphics[width=\textwidth]{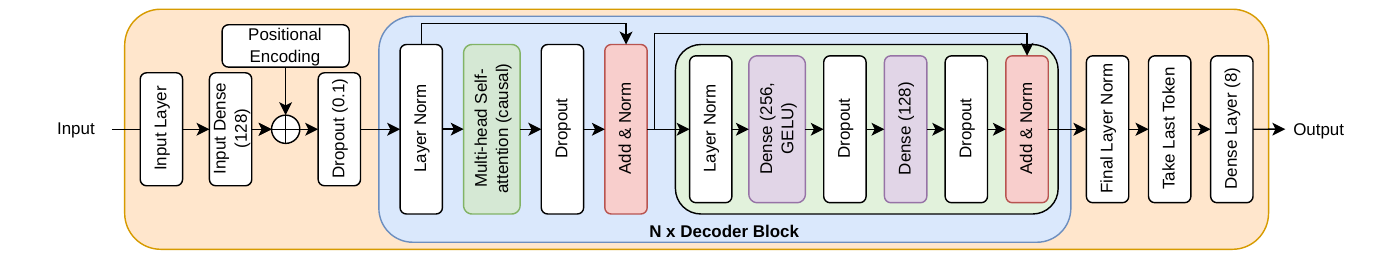}
    \caption{PAMPOS causal transformer-decoder architecture and parameters.}
    \label{fig:pampos}
\end{figure*}

\textbf{Data pre-processing.} Our experiments are conducted on the VeReMi++ dataset~\cite{kamel2020veremi}, which comprises 38 scenario subsets spanning two distinct 2-hour traffic periods: 19 morning and 19 afternoon subsets. The scenarios cover a broad behavioral spectrum, from benign vehicular mobility to common kinematic falsification attacks targeting position and speed, as well as \ac{DoS} scenarios. 

Each scenario contains two message types: type 2 messages originating from the ego vehicle (i.e., where PAMPOS is deployed) and type 3 messages corresponding to \acp{CAM} received from surrounding senders. Each received \ac{CAM} is aligned to the most recent ego state; messages whose temporal gap exceeds $2s$ are discarded to exclude delayed messages. The retained messages carry position, speed, acceleration, and heading, transformed into eight relative kinematic features as the delta between each \ac{CAM} field and the ego state, grouped by sender identity and sorted by receive time.

All messages from a given sender are treated as a single continuous sequence. Since vehicles periodically leave and re-enter the ego's communication range, the resulting sequences contain temporal gaps; we split sequences at these gaps to prevent abrupt feature discontinuities from contaminating training and inflating benign prediction errors. Feature-wise mean and standard deviation are computed from all benign sequences across the full 38-scenario corpus and applied for z-score normalization, ensuring a consistent input scale throughout training and evaluation.

\textbf{Training.} To prevent data leakage, sequences are partitioned into training, validation, and test sets by sender identity, ensuring that all windows belonging to a given sender reside exclusively within one split. Each sequence is then segmented using a sliding window of length 10 and stride 1, with the model trained to predict the 11th step. A practical challenge arises from the acceleration features ($\Delta\text{acl}_x$, $\Delta\text{acl}_y$), which exhibit occasional large-magnitude spikes in benign data due to sudden braking or sharp turning maneuvers. Under \ac{MSE}, a single outlier with a residual of 10 incurs a squared penalty of 100, disproportionately dominating the gradient signal and biasing the model toward over-predicting acceleration magnitude.

To mitigate this, we adopt the Huber loss~\cite{huber1964robust}, which interpolates between \ac{MSE} and \ac{MAE}: it preserves the differentiable quadratic regime for well-behaved predictions while bounding the gradient contribution of outliers through a linear tail. Formally, the per-sample, per-feature loss is defined as:
{
\begin{equation}
\footnotesize
    L_\delta(y, \hat{y}) = \begin{cases} \frac{1}{2}(y - \hat{y})^2 
    & \text{if } |y - \hat{y}| \leq \delta \\ 
    \delta\,|y - \hat{y}| - \frac{1}{2}\delta^2 & \text{otherwise,} 
    \end{cases}
\end{equation}
}
where $y$ is the true feature value, $\hat{y}$ is the model's prediction, and $\delta$ controls the transition between the quadratic and linear regimes. Setting $\delta = 1.0$ yields:

{
\begin{equation}
\footnotesize
    L(y, \hat{y}) = \begin{cases} \frac{1}{2}(y - \hat{y})^2 
    & \text{if } |y - \hat{y}| \leq 1 \\ 
    |y - \hat{y}| - \frac{1}{2} & \text{otherwise.} \end{cases}
\end{equation}
}
The total training objective is the mean Huber loss over all $N$ samples and $F = 8$ kinematic features:
{\begin{equation}
\footnotesize
   \mathcal{L}_{\text{total}} = \frac{1}{N} \sum_{n=1}^{N} \frac{1}{F} 
   \sum_{f=1}^{F} L\!\left(y_n^{(f)},\, \hat{y}_n^{(f)}\right).
\end{equation}
}

\begin{table}[!h]                                                                                   
  \centering                                                                                                  
  \caption{Transformer decoder architecture.}
  \label{tab:architecture}
  \footnotesize
  \begin{tabular}{l c}
      \toprule
      \textbf{Parameter} & \textbf{Value} \\
      \midrule
      Architecture style & Pre-norm (GPT-2) \\
      Hidden dimension ($d_\text{model}$) & 128 \\
      Attention heads & 8 (key dim = 16) \\
      Feed-forward dimension & 256 \\
      Decoder blocks & 3 \\
      Activation & GELU \\
      Dropout & 0.1 \\
      Positional encoding & Sinusoidal~\cite{vaswani2017attention} \\
      Total parameters & 399,880 \\
      \bottomrule
  \end{tabular}
\end{table}

We use a causal decoder-only design (Fig.~\ref{fig:pampos}) with 3 pre-norm blocks, 8 attention heads, and a hidden dimension of 128 (${\sim}400$K parameters), trained to predict the next trajectory step from a window of 10 consecutive relative-feature vectors. Each block follows the pre-norm layout, where layer normalization is applied before attention and feed-forward sublayers rather than after, providing more stable gradients during training. For positional encoding, we use the fixed sinusoidal functions of~\cite{vaswani2017attention}.

\textbf{Model Evaluation.} Our evaluation is twofold: first, we assess the ability to predict future values based on the sender's history, as smaller prediction errors enable better anomaly detection; second, we aggregate the top-K sender features with large errors to predict an attack using a threshold $\tau$. Specifically, we compute the per-feature absolute errors as:
{\begin{equation}
\footnotesize
    e_f = \left|y^{(f)} - \hat{y}^{(f)}\right| \quad f = 1, \ldots, 8
\end{equation}}

Then, we normalize by the benign \ac{MAE} of each feature, bringing all features to the same scale, enabling us to take the mean of the top-K largest errors as:
{\begin{equation}
\footnotesize
\label{eq:top-k}
    s = \frac{1}{K} \sum_{k=1}^{K} \tilde{e}_{(k)} \qquad \text{where } \tilde{e}_{(1)} \geq \tilde{e}_{(2)} \geq \cdots \geq \tilde{e}_{(8)}
\end{equation}
}
Selecting from the top subset of features enables us to detect attacks that affect only some of them. For sender-level detection, all window scores from a sender are averaged: 
{\begin{equation}
\label{eq:avg-attacker}
\footnotesize
    \bar{s}_{\text{sender}} = \frac{1}{|\mathcal{W}|} \sum_{w \in \mathcal{W}} s_w,
\end{equation}}
and a sender is flagged as an attacker if $\bar{s}_{\text{sender}} > \tau$. 
\section{Performance Evaluation}
\label{sec:performance}

\subsection{Setup}
\label{subsec:setup}

The transformer-decoder was implemented using TensorFlow~\cite{tf} and Keras~\cite{keras}. We performed our analysis on an Ubuntu machine with an AMD Ryzen Threadripper PRO 5965WX with 24 physical cores and 48 logical cores, 128 GB of available \ac{RAM}, and an NVIDIA GeForce RTX 4090 with 24 GB of DDR5 memory. For training the model, we use a batch size of 512, an initial \ac{LR} of $3 \times 10^{-4}$ for the Adam optimizer with gradient clipping at $\|\nabla\|_2 \leq 1.0$, and Huber loss ($\delta=1.0$) to reduce sensitivity to acceleration outliers. 

The \ac{LR} is halved after 4 epochs without improvement in validation loss (ReduceLROnPlateau), and training stops early after 8 stagnant epochs. The model is trained on the benign data from all 38 VeReMi++ scenarios (19 afternoon and 19 morning rush), yielding ${\sim}13$M training windows from 2,735 unique benign senders split 70/15/15 by sender identity to prevent data leakage. Table~\ref{tab:training-parameters} outlines all parameters used in the training and data setup.

The anomaly detection parameters used in this work are summarized in Table~\ref{tab:anomaly-parameters}. We compute the anomaly score for each incoming message based on the 3 largest per-feature normalized absolute errors (Eq.~\ref{eq:top-k}), and then average the window scores per sender (Eq.~\ref{eq:avg-attacker}) for vehicle-level classification. The threshold $\tau$ is set as a fixed percentile of the benign sender score distribution (Eq.~\ref{eq:avg-attacker}) using only benign validation data, requiring no attack labels; the chosen percentile directly controls the \ac{FPR}. During deployment, the operator selects the percentile based on their tolerable \ac{FPR}. For evaluation, we report results at the 98th percentile ($\tau_{98} = 4.43$), which corresponds to a 2\% \ac{FPR}; the test set is completely unseen by the model.

\begin{table}[!h]
      \centering
      \caption{Training and data parameters.}
      \label{tab:training-parameters}
      \resizebox{\columnwidth}{!}{
            \footnotesize
          \begin{tabular}{c c c c}
              \toprule
              \multicolumn{2}{c}{\textbf{Training}} & \multicolumn{2}{c}{\textbf{Data}} \\
              \cmidrule(lr){1-2} \cmidrule(lr){3-4}
              \textbf{Parameter} & \textbf{Value} & \textbf{Parameter} & \textbf{Value} \\
              \midrule
              Batch size & 512 & Scenarios & 38 (19 per period) \\
              \ac{LR} & $3 \times 10^{-4}$ & Attack types & 19 (A1--A19) \\
              Optimizer & Adam ($\|\nabla\|_2 \leq 1.0$) & Input features & 8 relative kinematics \\
              Loss function & Huber ($\delta = 1.0$) & Window size / stride & 10 / 1 \\
              LR schedule & ReduceLROnPlateau & Training windows & ${\sim}13$M \\
              LR reduce patience & 4 epochs & Data split (train/val/test) & 70/15/15 (by sender) \\
              Early stopping patience & 8 epochs & Min.\ sequence length & 15 messages \\
              & & Temporal gap threshold & 2$s$ \\
              \bottomrule
          \end{tabular}
      }
\end{table}

\begin{table}[!h]
      \centering
      \caption{Anomaly detection parameters.}
      \label{tab:anomaly-parameters}
      \footnotesize
      \begin{tabular}{l c}
          \toprule
          \textbf{Parameter} & \textbf{Value} \\
          \midrule
          Anomaly score & Mean-of-top-$K$ \\
          Top-$K$ features & 3 (of 8) \\
          Sender aggregation & Mean of window scores \\
          Deployment threshold ($\tau_{98}$) & 4.43 \\
          Threshold percentile & 98th \\
          Calibration set & 411 validation senders \\
          \bottomrule
      \end{tabular}
\end{table}

\subsection{Trajectory Prediction Analysis}
\label{sec:trajectory-prediction-analysis}
\begin{figure}[!t]
\centering
  \begin{subfigure}[h]{0.48\columnwidth}
   \includegraphics[width=\columnwidth]{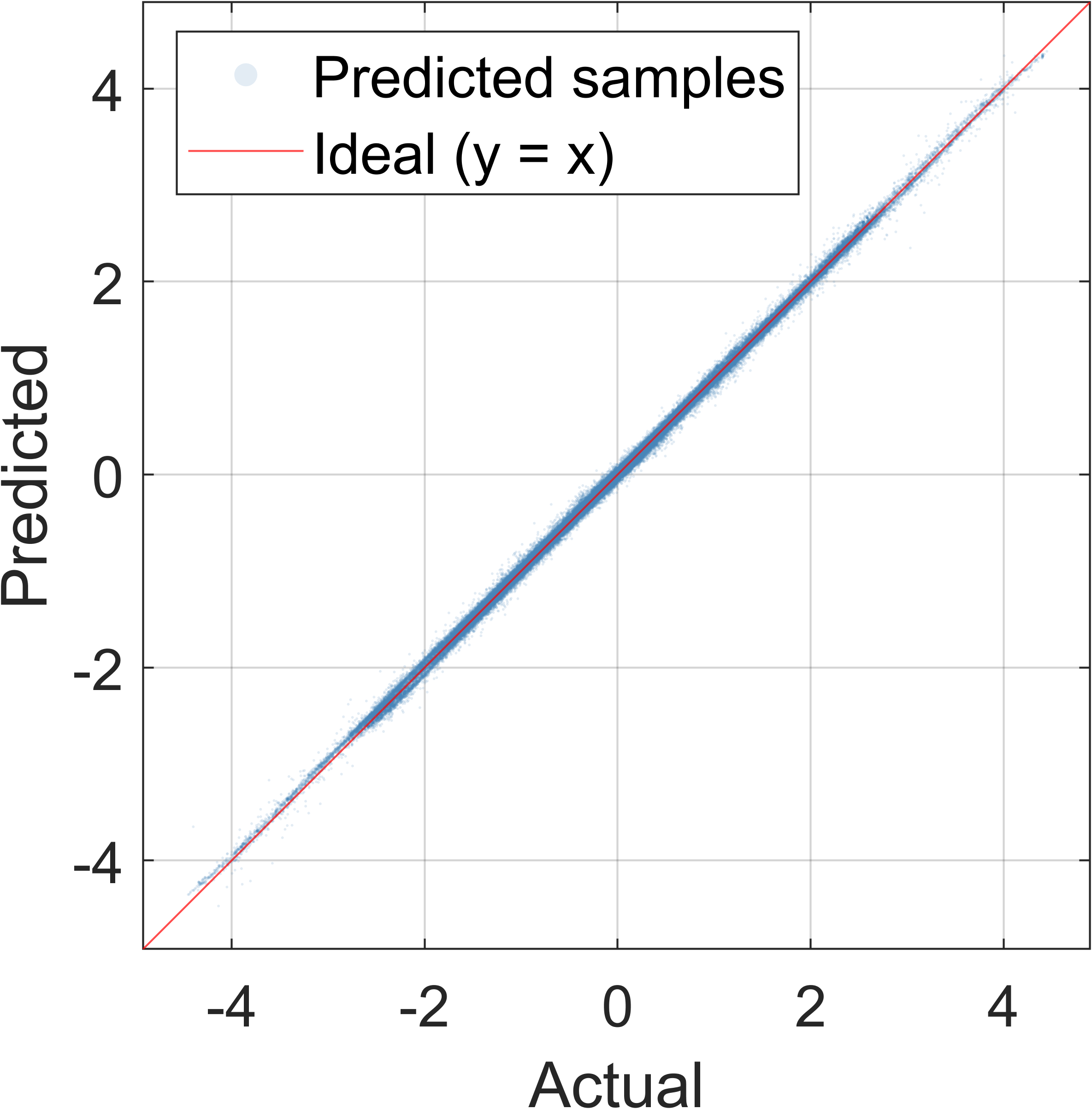}
    \caption{Position}
    \label{fig:scatter-position}
  \end{subfigure}
  \begin{subfigure}[h]{0.48\columnwidth}
    \includegraphics[width=\columnwidth]{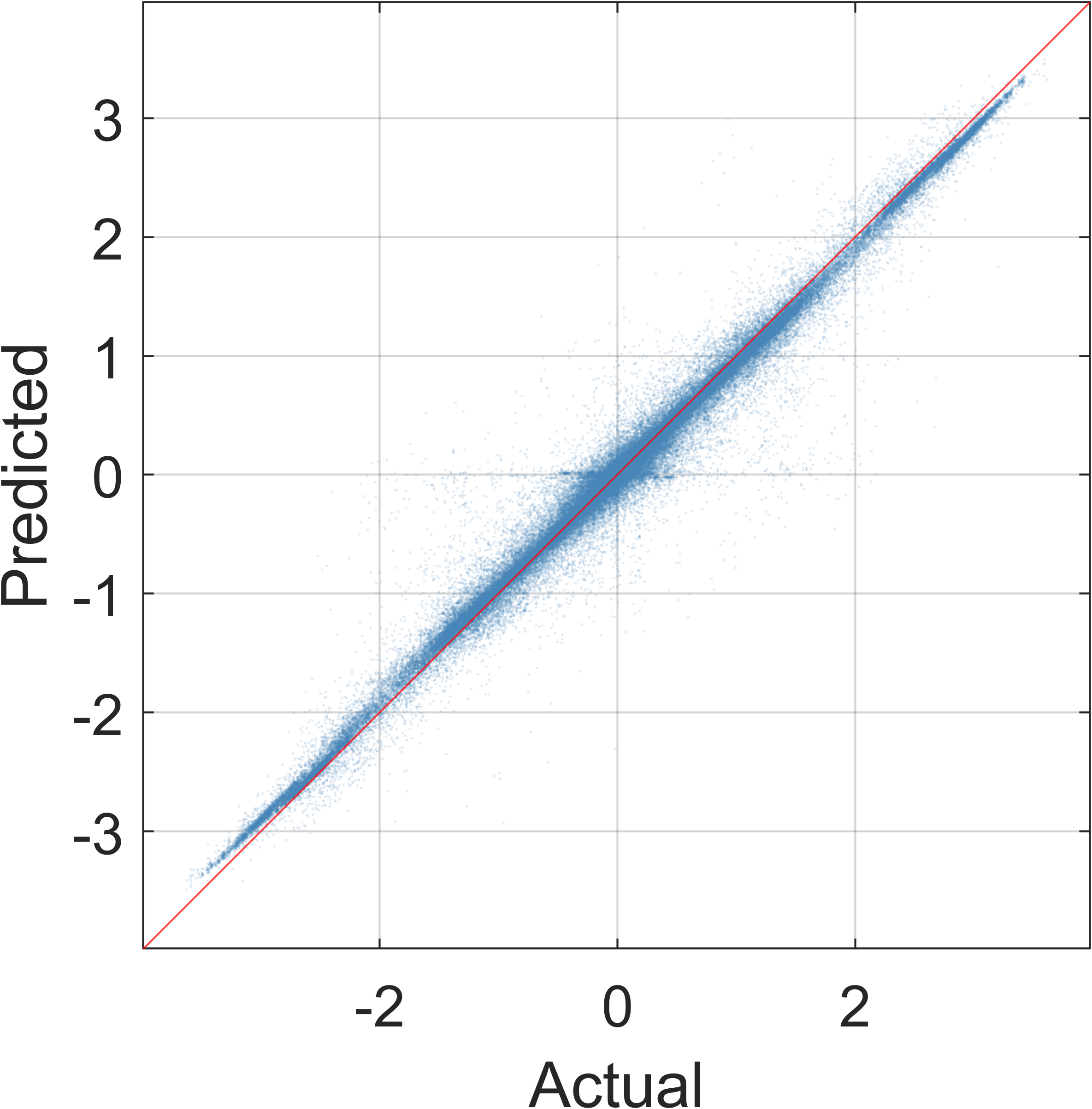}
    \caption{Speed}
    \label{fig:scatter-speed}
  \end{subfigure}
  \\
  \begin{subfigure}[h]{0.48\columnwidth}
   \includegraphics[width=\columnwidth]{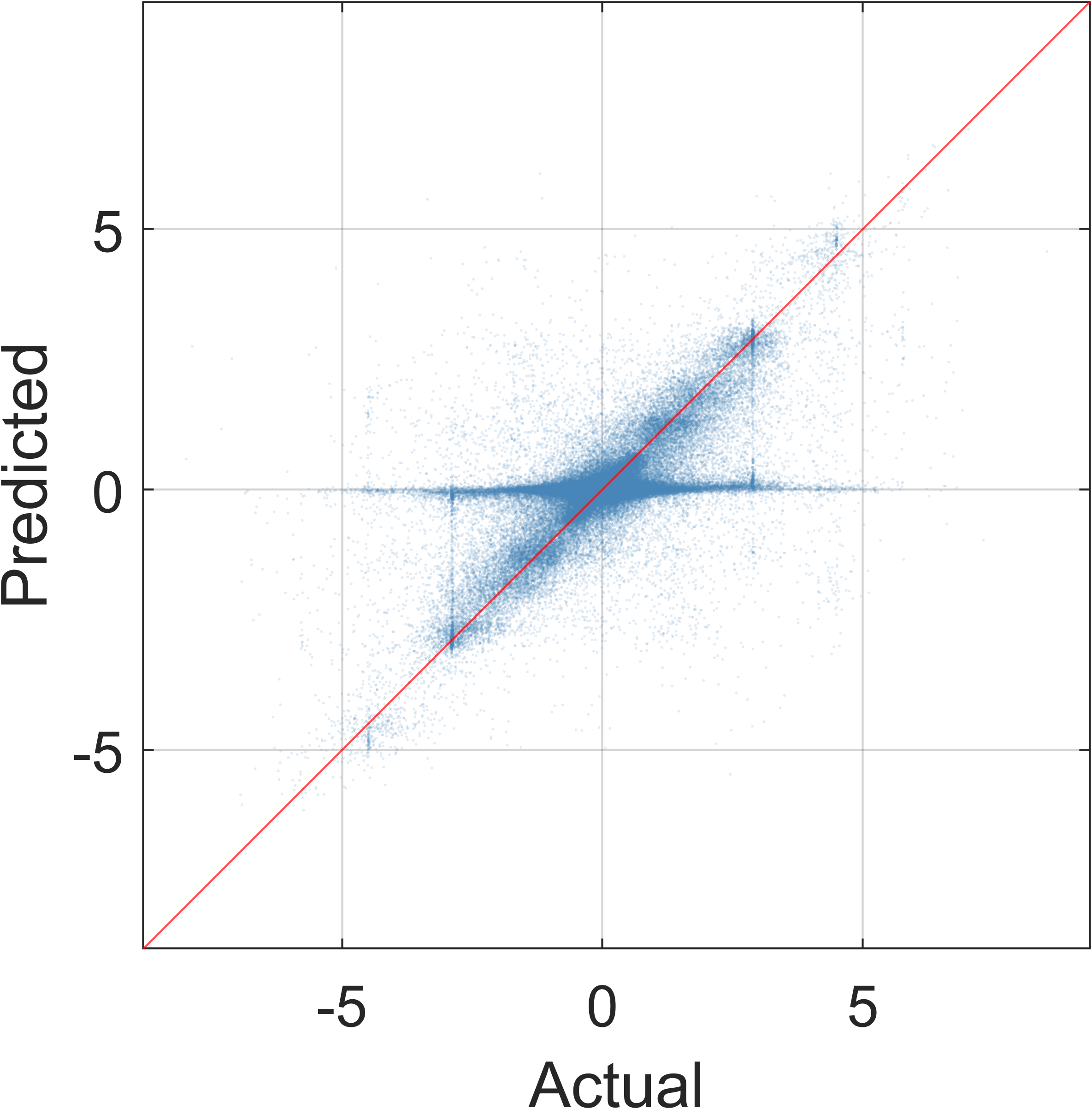}
    \caption{Acceleration}
    \label{fig:scatter-acceleration}
  \end{subfigure}
  \begin{subfigure}[h]{0.48\columnwidth}
    \includegraphics[width=\columnwidth]{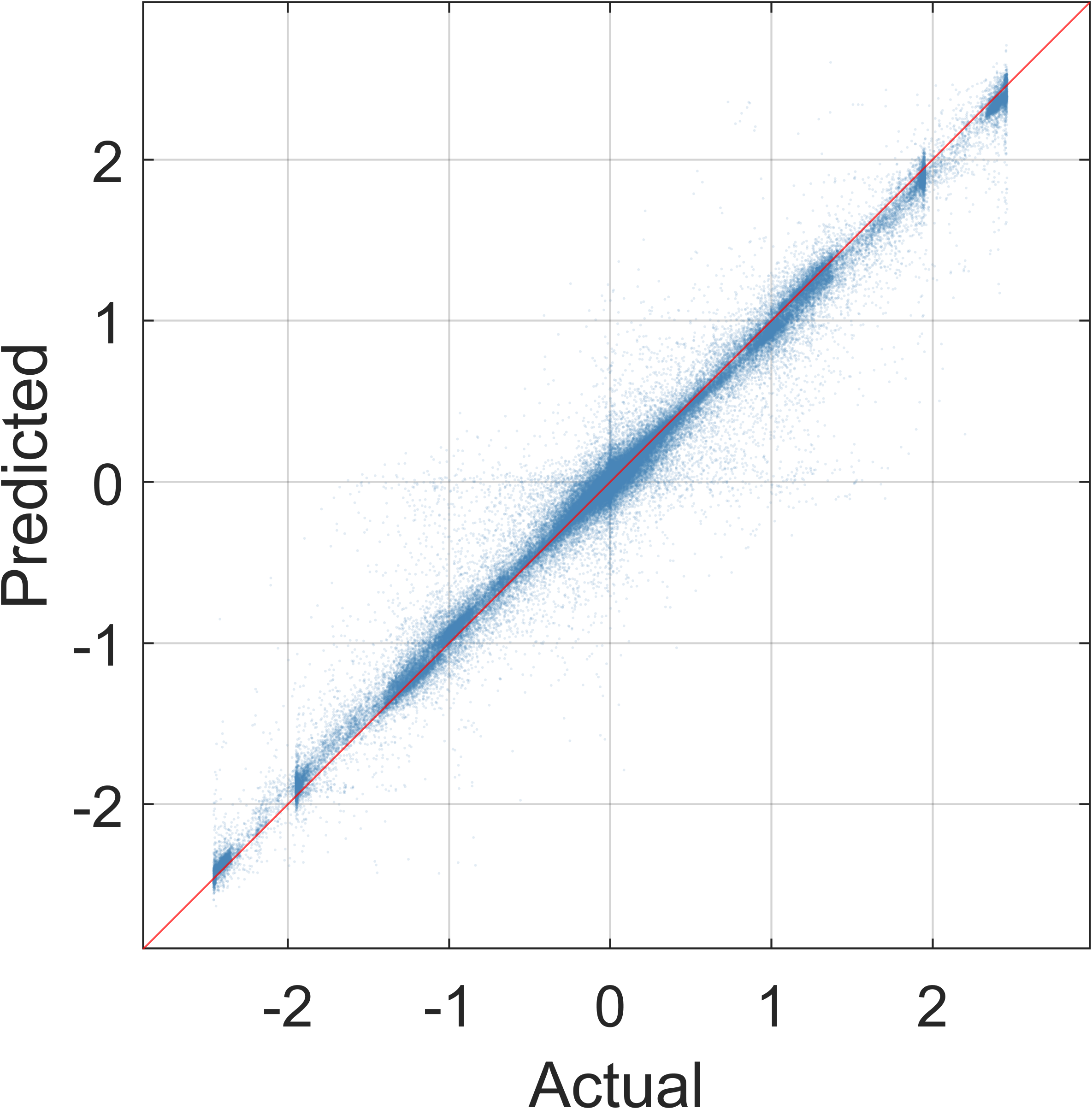}
    \caption{Heading}
    \label{fig:scatter-heading}
  \end{subfigure}
  \caption{Feature prediction error.}
  \label{fig:prediction-error}
\end{figure}

We assess the quality of transformer-decoder trajectory prediction, as the anomaly-scoring mechanism relies on the magnitudes of per-feature prediction residuals to discriminate between falsifying senders and benign ones. Fig.~\ref{fig:prediction-error} presents scatter plots of predicted vs. true values for each kinematic feature family, where the diagonal $y = x$ denotes perfect prediction. The model achieves alignment for position (Fig.~\ref{fig:scatter-position}), with moderate residuals in speed (Fig.~\ref{fig:scatter-speed}) and heading (Fig.~\ref{fig:scatter-heading}) attributable to abrupt lane changes. Acceleration (Fig.~\ref{fig:scatter-acceleration}) exhibits the largest errors, as it is the second derivative of position and thus inherently noisier, with sudden driver inputs producing large-magnitude changes that the model cannot anticipate from prior context alone. Overall, prediction fidelity is sufficient to yield meaningful separation between benign and attack residuals, as demonstrated in the following detection analysis.

\subsection{Misbehavior Detection Analysis}
\label{sec:mds-analysis}

\definecolor{heat-lowest}{HTML}{E74C3C}    
\definecolor{heat-low}{HTML}{F5B041}       
\definecolor{heat-medium-low}{HTML}{F7DC6F}
\definecolor{heat-medium}{HTML}{F9E79F}    
\definecolor{heat-medium-high}{HTML}{ABEBC6}
\definecolor{heat-high}{HTML}{82E0AA}      
\definecolor{heat-highest}{HTML}{2ECC71}   

\newcommand{\heatmapcell}[1]{%
  \ifdim#1pt>0.849pt
    \cellcolor{heat-lowest} #1
  \else\ifdim#1pt>0.699pt
    \cellcolor{heat-low} #1
  \else\ifdim#1pt>0.499pt
    \cellcolor{heat-medium-low} #1
  \else\ifdim#1pt>0.374pt
    \cellcolor{heat-medium} #1
  \else\ifdim#1pt>0.189pt
    \cellcolor{heat-medium-high} #1
  \else\ifdim#1pt>0.09pt
    \cellcolor{heat-high} #1
  \else
    \cellcolor{heat-highest} #1
  \fi\fi\fi\fi\fi\fi
}

{\color{customblue}
Table~\ref{tab:topk-frequency} shows the top-3 feature selection frequency across attack types. For each window, the anomaly scorer computes the per-feature normalized absolute error, i.e., the ratio of the prediction error to the benign baseline for that feature. Each cell in the table reports the fraction of windows where that feature appears among the selected top-3. i.e., if the selection were purely random, every feature would appear with frequency $3/8 = 0.375$. 

Benign senders (A0) approximate this uniform baseline (0.28 to 0.47), confirming that no single feature consistently dominates under normal conditions. In contrast, attack types show strong concentration on the features they distort. Position attacks (A1, A3, A4) select $\Delta x$ and $\Delta y$ in over 86\% of windows, while speed attacks (A7, A8) shift selection toward $\Delta \mathrm{spd}_x$ and $\Delta \mathrm{spd}_y$ (up to 0.79). Notably, acceleration features ($\Delta \mathrm{acl}$) are rarely selected (as low as 0.00 for A14 and A18) because position or speed attacks produce errors on the other features that are larger than acceleration noise. 

Specifically, A2 (Constant Position Offset) exhibits near-uniform selection that closely resembles benign traffic. Its anomaly signal is diffuse rather than concentrated in specific features, as the small constant offset keeps the falsified trajectory behaviorally consistent with normal driving. Choosing a different value of K would not improve the detector. On the other hand, A9 (Eventual Stop), A13 (\ac{DoS}), A16 (Grid Sybil), and A17 (Data Replay Sybil) show moderate position-feature concentration (0.70--0.79 for $\Delta x$, $\Delta y$), which could suggest $K{=}2$; conversely, speed attacks (A7, A8) would suggest $K{=}4$, though this would dilute the scores for position-concentrated attacks (A1, A3). $K{=}3$ provides a conservative, but robust default. Nonetheless, top-$K$ co-selection patterns offer insights into the attack family: speed co-selection identifies speed attacks (A5, A7, A8), while heading co-selection identifies disruptive attacks.
}
\begin{table}[!h] 
  \centering
  \caption{Top-3 feature selection frequency per attack type.}                                                                                                                                                                                                                                 
  \label{tab:topk-frequency}
  \small
  \resizebox{0.99\columnwidth}{!}{%
      \setlength{\tabcolsep}{3.5pt}
      \begin{tabular}{clcccccccc}
    \toprule
     & & $\Delta x$ & $\Delta y$ & $\Delta \mathrm{spd}_x$ & $\Delta \mathrm{spd}_y$ & $\Delta \mathrm{acl}_x$ & $\Delta \mathrm{acl}_y$ & $\Delta \mathrm{hed}_x$ & $\Delta \mathrm{hed}_y$ \\
    \midrule
    \textbf{A0} & Benign & \heatmapcell{0.43} & \heatmapcell{0.47} & \heatmapcell{0.32} & \heatmapcell{0.41} & \heatmapcell{0.28} & \heatmapcell{0.35} & \heatmapcell{0.34} & \heatmapcell{0.40} \\
    \midrule
    \textbf{A1} & Const.\ Position & \heatmapcell{0.94} & \heatmapcell{0.93} & \heatmapcell{0.14} & \heatmapcell{0.20} & \heatmapcell{0.13} & \heatmapcell{0.10} & \heatmapcell{0.26} & \heatmapcell{0.30} \\
    \textbf{A2} & Const.\ Pos.\ Offset & \heatmapcell{0.47} & \heatmapcell{0.57} & \heatmapcell{0.26} & \heatmapcell{0.45} & \heatmapcell{0.23} & \heatmapcell{0.28} & \heatmapcell{0.26} & \heatmapcell{0.49} \\
    \textbf{A3} & Random Position & \heatmapcell{0.98} & \heatmapcell{0.97} & \heatmapcell{0.17} & \heatmapcell{0.22} & \heatmapcell{0.10} & \heatmapcell{0.11} & \heatmapcell{0.25} & \heatmapcell{0.20} \\
    \textbf{A4} & Random Pos.\ Offset & \heatmapcell{0.91} & \heatmapcell{0.86} & \heatmapcell{0.17} & \heatmapcell{0.19} & \heatmapcell{0.21} & \heatmapcell{0.25} & \heatmapcell{0.18} & \heatmapcell{0.23} \\
    \midrule
    \textbf{A5} & Const.\ Speed & \heatmapcell{0.85} & \heatmapcell{0.80} & \heatmapcell{0.53} & \heatmapcell{0.40} & \heatmapcell{0.06} & \heatmapcell{0.04} & \heatmapcell{0.16} & \heatmapcell{0.17} \\
    \textbf{A6} & Const.\ Speed Offset & \heatmapcell{0.96} & \heatmapcell{0.84} & \heatmapcell{0.29} & \heatmapcell{0.20} & \heatmapcell{0.20} & \heatmapcell{0.14} & \heatmapcell{0.17} & \heatmapcell{0.19} \\
    \textbf{A7} & Random Speed & \heatmapcell{0.79} & \heatmapcell{0.60} & \heatmapcell{0.77} & \heatmapcell{0.64} & \heatmapcell{0.02} & \heatmapcell{0.02} & \heatmapcell{0.08} & \heatmapcell{0.08} \\
    \textbf{A8} & Random Speed Offset & \heatmapcell{0.68} & \heatmapcell{0.46} & \heatmapcell{0.79} & \heatmapcell{0.70} & \heatmapcell{0.09} & \heatmapcell{0.07} & \heatmapcell{0.12} & \heatmapcell{0.09} \\
    \midrule
    \textbf{A9} & Eventual Stop & \heatmapcell{0.70} & \heatmapcell{0.74} & \heatmapcell{0.23} & \heatmapcell{0.30} & \heatmapcell{0.13} & \heatmapcell{0.18} & \heatmapcell{0.35} & \heatmapcell{0.38} \\
    \textbf{A10} & Disruptive & \heatmapcell{0.79} & \heatmapcell{0.80} & \heatmapcell{0.25} & \heatmapcell{0.30} & \heatmapcell{0.02} & \heatmapcell{0.06} & \heatmapcell{0.40} & \heatmapcell{0.39} \\
    \textbf{A11} & Data Replay & \heatmapcell{0.73} & \heatmapcell{0.76} & \heatmapcell{0.26} & \heatmapcell{0.31} & \heatmapcell{0.09} & \heatmapcell{0.15} & \heatmapcell{0.36} & \heatmapcell{0.34} \\
    \textbf{A12} & Delayed Messages & \heatmapcell{0.85} & \heatmapcell{0.91} & \heatmapcell{0.16} & \heatmapcell{0.20} & \heatmapcell{0.07} & \heatmapcell{0.12} & \heatmapcell{0.27} & \heatmapcell{0.41} \\
    \midrule
    \textbf{A13} & DoS & \heatmapcell{0.71} & \heatmapcell{0.76} & \heatmapcell{0.22} & \heatmapcell{0.28} & \heatmapcell{0.28} & \heatmapcell{0.28} & \heatmapcell{0.20} & \heatmapcell{0.25} \\
    \textbf{A14} & DoS Random & \heatmapcell{0.95} & \heatmapcell{0.91} & \heatmapcell{0.53} & \heatmapcell{0.29} & \heatmapcell{0.00} & \heatmapcell{0.00} & \heatmapcell{0.20} & \heatmapcell{0.11} \\
    \textbf{A15} & DoS Disruptive & \heatmapcell{0.79} & \heatmapcell{0.81} & \heatmapcell{0.27} & \heatmapcell{0.28} & \heatmapcell{0.03} & \heatmapcell{0.04} & \heatmapcell{0.41} & \heatmapcell{0.38} \\
    \midrule
    \textbf{A16} & Grid Sybil & \heatmapcell{0.74} & \heatmapcell{0.71} & \heatmapcell{0.32} & \heatmapcell{0.34} & \heatmapcell{0.15} & \heatmapcell{0.15} & \heatmapcell{0.33} & \heatmapcell{0.25} \\
    \textbf{A17} & Data Replay Sybil & \heatmapcell{0.74} & \heatmapcell{0.79} & \heatmapcell{0.24} & \heatmapcell{0.31} & \heatmapcell{0.10} & \heatmapcell{0.13} & \heatmapcell{0.35} & \heatmapcell{0.34} \\
    \textbf{A18} & DoS Random Sybil & \heatmapcell{0.95} & \heatmapcell{0.91} & \heatmapcell{0.49} & \heatmapcell{0.29} & \heatmapcell{0.00} & \heatmapcell{0.00} & \heatmapcell{0.21} & \heatmapcell{0.14} \\
    \textbf{A19} & DoS Disruptive Sybil & \heatmapcell{0.86} & \heatmapcell{0.85} & \heatmapcell{0.24} & \heatmapcell{0.22} & \heatmapcell{0.03} & \heatmapcell{0.03} & \heatmapcell{0.42} & \heatmapcell{0.34} \\
    \bottomrule
    \end{tabular}
   }
\end{table}

\begin{figure}[!htbp]
    \centering
    \includegraphics[width=0.75\linewidth]{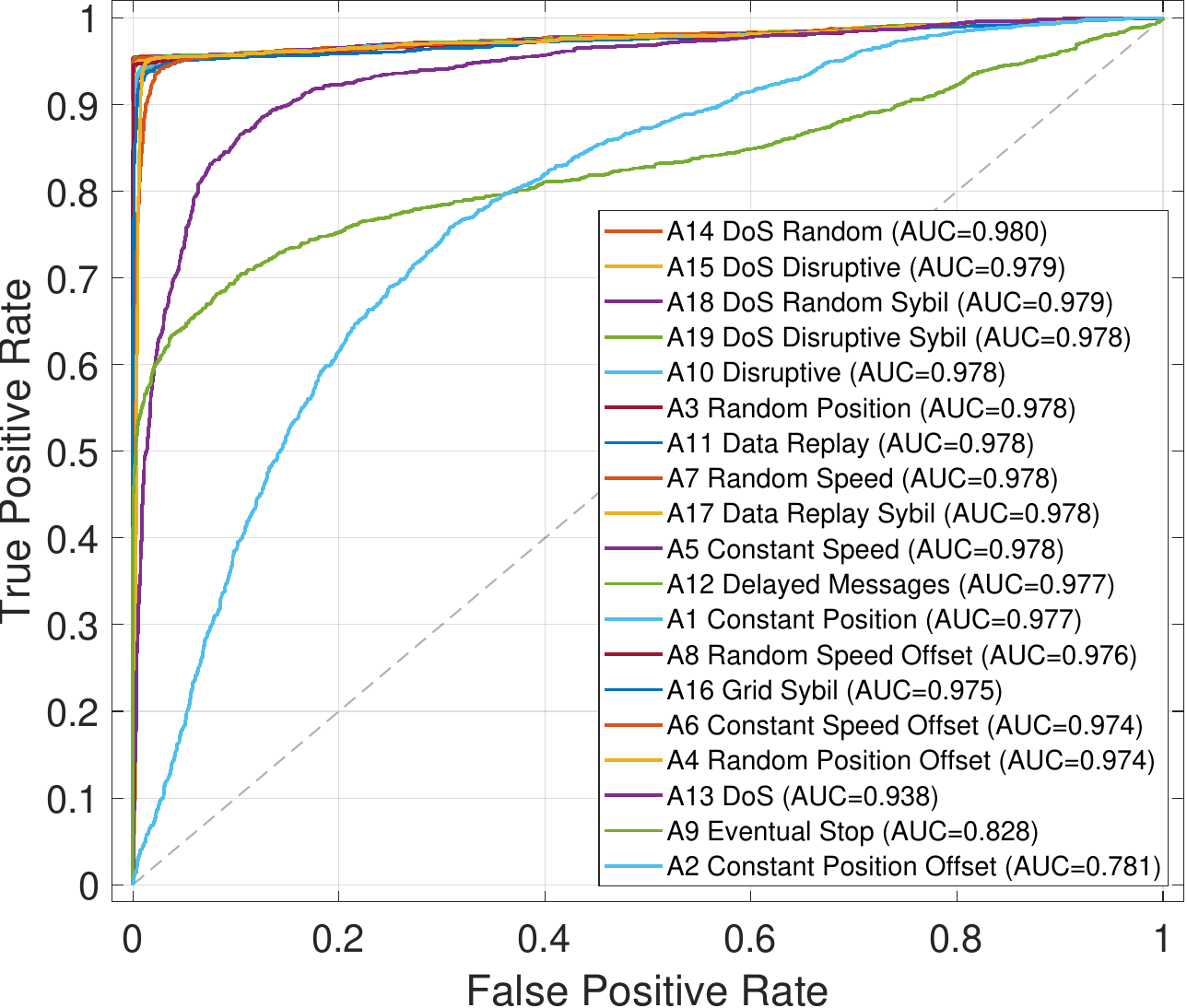}
    \caption{\acs{ROC} curves for each attack scenario.}
    \label{fig:roc-attacks}
\end{figure}

Fig.~\ref{fig:roc-attacks} presents the \ac{ROC} curves for all nineteen attack scenarios, with the legend sorted by \ac{AUC}. The majority of attacks achieve \ac{AUC} values between 0.93 and 0.98, with consistently high \ac{TPR} at low \ac{FPR}. The notable exceptions are A13, A9, and A2, which exhibit progressively degraded performance. A9 and A2 display qualitatively distinct curve characteristics: A9 maintains a higher \ac{TPR} than A2 up to a \ac{FPR} of 0.38, beyond which the performance reverses.

\newcommand{\metricscell}[1]{%
  \ifdim#1pt<0.83pt
    \cellcolor{heat-lowest} #1
  \else\ifdim#1pt<0.86pt
    \cellcolor{heat-low} #1
  \else\ifdim#1pt<0.89pt
    \cellcolor{heat-medium-low} #1
  \else\ifdim#1pt<0.92pt
    \cellcolor{heat-medium} #1
  \else\ifdim#1pt<0.95pt
    \cellcolor{heat-medium-high} #1
  \else\ifdim#1pt<0.98pt
    \cellcolor{heat-high} #1
  \else
    \cellcolor{heat-highest} #1
  \fi\fi\fi\fi\fi\fi
}

\begin{table*}[!ht]
\centering
\caption{Attack detection metrics for each attack category.}
\label{tab:metrics}
\resizebox{0.99\textwidth}{!}{%
    \begin{tabular}{c | c c c c | c c c c | c | c c c | c c c | c c c c }
        \toprule
        \textbf{Metric} & \multicolumn{4}{c}{\textbf{Position}} & \multicolumn{4}{c}{\textbf{Speed}} & \multicolumn{1}{c}{\textbf{Eventual Stop}} & \multicolumn{3}{c}{\textbf{Other}} & \multicolumn{3}{c}{\textbf{DoS}} & \multicolumn{4}{c}{\textbf{Sybil}} \\
        \cmidrule(lr){2-5} \cmidrule(lr){6-9} \cmidrule(lr){10-10} \cmidrule(lr){11-13} \cmidrule(lr){14-16} \cmidrule(lr){17-20}
        & A1 & A2 & A3 & A4 & A5 & A6 & A7 & A8 & A9 & A10 & A11 & A12 & A13 & A14 & A15 & A16 & A17 & A18 & A19 \\
        \hline
        Accuracy  & \metricscell{0.97} & \metricscell{0.69} & \metricscell{0.97} & \metricscell{0.97} & \metricscell{0.97} & \metricscell{0.96} & \metricscell{0.97} & \metricscell{0.97} & \metricscell{0.86} & \metricscell{0.97} & \metricscell{0.97} & \metricscell{0.97} & \metricscell{0.86} & \metricscell{0.97} & \metricscell{0.97} & \metricscell{0.96} & \metricscell{0.97} & \metricscell{0.97} & \metricscell{0.97} \\
        Precision & \metricscell{0.94} & \metricscell{0.63} & \metricscell{0.95} & \metricscell{0.94} & \metricscell{0.94} & \metricscell{0.94} & \metricscell{0.95} & \metricscell{0.95} & \metricscell{0.92} & \metricscell{0.95} & \metricscell{0.95} & \metricscell{0.94} & \metricscell{0.92} & \metricscell{0.94} & \metricscell{0.94} & \metricscell{0.94} & \metricscell{0.95} & \metricscell{0.94} & \metricscell{0.94} \\
        Recall    & \metricscell{0.95} & \metricscell{0.11} & \metricscell{0.95} & \metricscell{0.95} & \metricscell{0.95} & \metricscell{0.94} & \metricscell{0.95} & \metricscell{0.95} & \metricscell{0.61} & \metricscell{0.95} & \metricscell{0.95} & \metricscell{0.95} & \metricscell{0.64} & \metricscell{0.96} & \metricscell{0.96} & \metricscell{0.95} & \metricscell{0.95} & \metricscell{0.96} & \metricscell{0.96} \\
        F1        & \metricscell{0.95} & \metricscell{0.19} & \metricscell{0.95} & \metricscell{0.95} & \metricscell{0.95} & \metricscell{0.94} & \metricscell{0.95} & \metricscell{0.95} & \metricscell{0.73} & \metricscell{0.95} & \metricscell{0.95} & \metricscell{0.95} & \metricscell{0.75} & \metricscell{0.95} & \metricscell{0.95} & \metricscell{0.94} & \metricscell{0.95} & \metricscell{0.95} & \metricscell{0.95} \\
        \bottomrule
    \end{tabular}
}
\end{table*}

The overall detection metrics are reported in Table~\ref{tab:metrics}. \textcolor{customblue}{\ac{DoS} attacks A14 to A19 are detected despite the absence of explicit temporal features because their \acp{CAM} carry replayed or kinematically altered data, producing large prediction residuals regardless of transmission frequency. The simple \ac{DoS} (A13) is a partial exception: its messages carry valid kinematics with only the frequency altered, compressing the 10-message window to milliseconds rather than the expected ${\sim}1$\,s. The resulting near-zero per-step deltas, which the model expects for normal speeds, yield a moderate anomaly signal (F1\,=\,0.75), confirming that the learned temporal structure implicitly encodes expected transmission intervals.}

\begin{figure}[!h]
\centering
  \begin{subfigure}[h]{0.49\columnwidth}
   \includegraphics[width=\columnwidth]{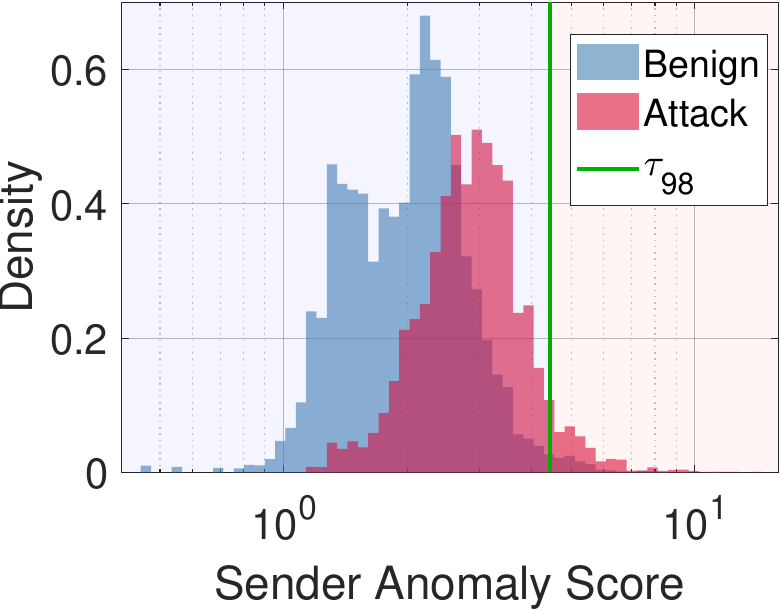}
    \caption{Const. Position Offset}
    \label{fig:a2}
  \end{subfigure}
  \begin{subfigure}[h]{0.49\columnwidth}
    \includegraphics[width=\columnwidth]{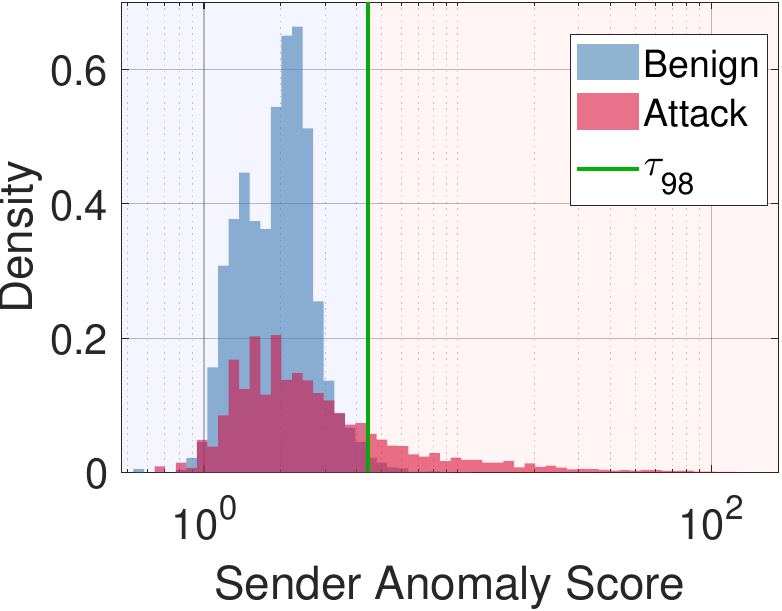}
    \caption{Eventual Stop}
    \label{fig:a9}
  \end{subfigure}
  \caption{Anomaly scores.}
  \label{fig:thresholds}
\end{figure}

The most challenging cases are A2 (Constant Position Offset, F1 = 0.19) and A9 (Eventual Stop, F1 = 0.73). \textcolor{customblue}{A2 small constant shift preserves all higher-order kinematics (speed, acceleration, heading) and trajectory shape, keeping the anomaly score distribution overlapping with the benign one (Fig.~\ref{fig:thresholds}), while the top-$3$ selection frequencies (0.23 to 0.57) are indistinguishable from the benign baseline (0.28 to 0.47, Table~\ref{tab:topk-frequency}), leaving no discriminative feature.} A9 performs better, though \textcolor{customblue}{early deceleration is indistinguishable from normal braking; the anomaly signal emerges only once the vehicle remains stationary.} Despite this, our solution has comparable overall F1 scores with supervised solutions that require labeled data during training (recall Sec.~\ref{sec:background}).

\subsection{Cross-Scenario Generalization}
\label{sec:discussion}

To test generalization, we train a model exclusively on afternoon (14:00 to 16:00) benign data and evaluate on morning rush hour (07:00 to 09:00) attacks, featuring different traffic densities and vehicle populations. The threshold $\tau_{96}$ (producing the optimal F1 score) is selected with the same method, as described in Sec.~\ref{subsec:setup}. \textcolor{customblue}{The afternoon-trained model achieves a mean AUC of 0.95 and a mean F1 of 0.87 across all 19 attacks. 16 of those retain F1 $\geq$ 0.91, with the largest drops confined to A2 (F1 = 0.14), A9 (F1 = 0.70), and A13 (F1 = 0.74). Training in the reverse direction (morning rush to afternoon) yields nearly identical aggregate performance (mean \ac{AUC} = 0.95, mean F1 = 0.86), confirming symmetric generalization. This is expected: relative kinematic features are invariant to absolute position, road topology, and traffic density, so patterns learned from afternoon traffic transfer directly to rush hour conditions.}

\section{Conclusion}
\label{sec:conclusion}
We presented \acs{PAMPOS}, an unsupervised \ac{MDS} for \ac{V2X} networks based on a causal transformer-decoder trained only on benign trajectories from VeReMi++. \acs{PAMPOS} identifies falsifying senders as those whose reported kinematic sequences deviate significantly from a learned mobility prior, without requiring any attack-labeled training data. Evaluated across all 19 attack types in VeReMi++, \acs{PAMPOS} achieves \ac{AUC} values of up to 0.98 and F1-scores of up to 0.95 for the majority of attack categories. The top-$K$ normalized scoring mechanism successfully localizes falsification to the specific kinematic features each attack distorts, enabling detection of attacks that corrupt only a subset of the reported state. Bidirectional cross-scenario evaluation confirms symmetric generalization between afternoon and morning rush hour traffic, owing to the scenario-invariant relative features. The constant position offset (A2) and eventual stop (A9) attacks remain open challenges, as their falsified trajectories are behaviorally indistinguishable from benign mobility under the learned model; addressing these cases requires complementary strategies such as cross-sender consistency checks or map-aware validation. Future work will investigate augmented scoring strategies to address these hard cases, an ablation study on all parameters, incorporate temporal frequency features to strengthen \ac{DoS} detection, and explore model compression for deployment on resource-constrained devices.

\begin{acks}
This work is supported in parts by the Swedish Research Council (VR) and the Knut and Alice Wallenberg (KAW) Foundation.
\end{acks}

\bibliographystyle{ACM-Reference-Format}
\bibliography{main}

\end{document}